\documentclass[aps,prd,twocolumn,showpacs,amsmath,amssymb,nofootinbib,eqsecnum, preprintnumbers]{revtex4-1}
 
\usepackage{hyperref} 
\usepackage{graphicx}
\usepackage{epstopdf}

\newcommand{\be}{\begin{equation}}
\newcommand{\ee}{\end{equation}}
\newcommand{\ba}{\begin{eqnarray}}
\newcommand{\ea}{\end{eqnarray}}
\newcommand{\nn}{\nonumber} 
\newcommand{\cH}{\mathcal{H}} 
\newcommand{\cL}{\mathcal{L}} 
\newcommand{\dl}[1]{\frac{d #1}{d \lambda}} 
\newcommand{\dt}[1]{\frac{d #1}{d t}}

\begin{document}
\title{The Eisenhart lift: a didactical introduction of modern geometrical concepts from Hamiltonian dynamics}

\author{Marco Cariglia}
\email{marco@iceb.ufop.br}
\affiliation{Universidade Federal de Ouro Preto, ICEB, Departamento de F\'isica.
  Campus Morro do Cruzeiro, Morro do Cruzeiro, 35400-000 - Ouro Preto, MG - Brasil}
 
\author{Filipe Kelmer Alves} 
\email{fkelmer9@gmail.com} 
\affiliation{Universidade Federal de Ouro Preto, ICEB, Departamento de F\'isica.
  Campus Morro do Cruzeiro, Morro do Cruzeiro, 35400-000 - Ouro Preto, MG - Brasil}

\date{\today}  

\begin{abstract} 
This work originates from part of a final year undergraduate research project on the Eisenhart lift for Hamiltonian systems. The Eisenhart lift is a procedure to describe trajectories of a classical natural Hamiltonian system as geodesics in an enlarged space. We point out that it can be easily obtained from basic principles of Hamiltonian dynamics, and as such it represents a useful didactical way to introduce graduate students to several modern concepts of geometry applied to physics: curved spaces, both Riemannian and Lorentzian, conformal transformations, geometrisation of interactions and extra dimensions, geometrisation of dynamical symmetries. For all these concepts the Eisenhart lift can be used as a theoretical tool that provides easily achievable examples, with the added benefit of also being a topic of current research with several applications, among which the study of dynamical systems and non-relativistic holography. 
\end{abstract}

\maketitle

\section{Introduction} 
Geometry and geometrical methods have always played an important role in the study of physics and mathematics. In particular in the last and in the current century we have witnessed an increase in the range of what geometry means and what kind of physical questions it can help tackle. As a chief example, but not the only one, we have in mind the bigger role taken on by geometry in describing interactions in theoretical physics. From the early work of Einstein describing gravity in terms of a curved spacetime, to the seminal idea of Kaluza and Klein of including other interactions by adding extra dimensions \cite{Kaluza,Klein}, we have been lead into the framework of modern theoretical approaches like String Theory or braneworld cosmologies where extra dimensions and their dynamics are directly dictating the physical laws and the structure of possible fields. Such innovative ideas are however, by nature of their higher mathematical content, typically taught at Universities only at the postgraduate level in theoretical oriented courses. Undergraduate students and non-theoretical ones will likely miss out. However, this needs not be the case, since it is possible to introduce a number of modern geometrical concepts using the simpler setting of the Eisenhart lift of natural dynamical systems. 
 
The lift procedure was introduced by Eisenhart at the beginning of last century \cite{Eisenhart1928}, and provides an interpretation of trajectories of natural Hamiltonian systems in terms of geodesics in a curved space. This concept is not completely new, a related approach is that of the Jacobi metric. However, differently from the Jacobi metric the Eisenhart lift has the specific property of requiring to enlarge the number of dimensions of the system studied: geometrisation of interactions is achieved introducing one or more extra dimensions. It should be noticed here that the geometrisation achieved is far simpler in type and scope than that of String Theory or other related theories, however in the present context of a didactical introduction we find this a point of advantage rather than a weakness: it is easy to see the main reasons why having extra dimensions can be a good thing with relatively little pre-requisites. Similarly, one can naturally introduce the concepts of curved metrics, both Riemannian and Lorentzian, and that of conformal transformations: the Lorentzian lift is done in terms of null geodesics, and the student can see an example of the application of conformal transformations by seeing explicitly how the conformal group of flat Lorentzian space is related to the group of symmetries of dynamics for the free particle. More generally, the student can see a relation between symmetries of dynamics for the original system and conformal transformations for the lift geometry. This is at the foundation of modern research on non-relativistic holography and, while it will be the teacher's decision on how much detail to include in the presentation, it is possible to guide the student into exploring this kind of relationships that are proving so fruitful in  moden research. 
 
It is the aim of this work to give a pedagogical introduction to the Eisenhart lift, showing how it can be justified from standard concepts of Hamiltonian dynamics and with little pre-requisite other than knowledge of the content of a standard undergraduate 'rational mechanics'  course. Some discussion of curved spaces will be necessary, and its detail can be gauged by the teacher.  We will also give a brief account of modern research based on the Eisenhart lift framework: notable examples are the study of integrability and chaos in dynamical systems and non-relativistic holography. The audience we have in mind is that of undergraduate and master students with a good understanding of Hamiltonian dynamics, as well as higher education teachers.

\section{From Hamiltonian dynamics to the geometrical lift} 
 
\subsection{The Riemannian lift} 
Our starting point will be that of a \textit{natural Hamiltonian system}, i.e. a dynamical system whose Hamiltonian is a quadratic function of the momenta. In the most general case, this includes both a scalar and a vector potential, the last one associated to a magnetic field. For the purpose of simplicity we will begin with just the scalar potential, and add the vector potential later. In terms of $n$ generalised position coordinates $q^i$ and $n$ momenta $p_i$ our Hamiltonian therefore will be 
\be \label{eq:Hamiltonian_scalar} 
H = \frac{1}{2} \sum_{i,j = 1}^n h^{ij}(q) p_i p_j + e^2 V(q) \, . 
\ee 
Here, $h^{ij}(q) = h^{ji}(q)$ is a symmetric matrix function of the $q$ coordinates,  $V(q)$ is the scalar potential and $e$ a constant. We will ask $h^{ij}$ to be invertible so that there exists a one to one relation between momenta and velocities calculated in the standard way as $\dt{q^i} = \sum_i h^{ij} p_j$, and denote the inverse matrix by $h_{ij} = h_{ji}$. We will also ask that $h^{ij}$ is strictly positive definite so that the first term in \eqref{eq:Hamiltonian_scalar} represents a positive kinetic energy. 
A student acquainted with curved geometry will recognise that $h$ amounts to a Riemannian metric on configuration space, but in general it will be sufficient to notice that this represents the most general way to obtain a positive  position dependent quadratic function of the momenta. The equations of motion are: 
\ba 
\dt{q^i} &=& \frac{\partial H}{\partial p_i} = \sum_i h^{ij} p_j \, , \nn \\ 
\dt{p_i} &=& - \frac{\partial H}{\partial q^i} = - \frac{1}{2} \partial_i h^{jk} p_j p_k - e^2 \partial_i V  \, . \nn 
\ea 
The main idea in the Eisenhart lift construction is noticing that we can make $H$ a \textit{homogeneous} quadratic function of the momenta. More precisely we can consider a new function that depends on $q^i$, $p_j$ and a new momentum $p_y$ 
\be 
\cH = \frac{1}{2} \sum_{i,j = 1}^n h^{ij}(q) p_i p_j + p_y^2 V(q)  \, . 
\ee 
This will reduce to the original Hamiltonian \eqref{eq:Hamiltonian_scalar} when $p_y = q$. There is nothing prohibiting us to think of $\cH$ as the Hamiltonian of a new system with additional conjugate variables $y$ and $p_y$: the new equations of motion will be 
\ba 
\dt{q^i} &=& \frac{\partial \cH}{\partial p_i} = \sum_i h^{ij} p_j \, , \label{eq:Eisenhart_scalar_eom1}  \\ 
\dt{p_i} &=& - \frac{\partial \cH}{\partial q^i} =  - \frac{1}{2} \partial_i h^{jk} p_j p_k - p_y^2 \partial_i V  \, , \label{eq:Eisenhart_scalar_eom2}  \\ 
\dt{y} &=& \frac{\partial \cH}{\partial p_y} = 2 p_y V(q) \, , \label{eq:Eisenhart_scalar_eom3}  \\ 
\dt{p_y} &=& \frac{\partial \cH}{\partial y} = 0 \, ,  \label{eq:Eisenhart_scalar_eom4} 
\ea 
The last equation guarantees that we can take $p_y = e$ and therefore recover the equations of motion of the original system; the equation for $y$ can be solved once the trajectory $t \mapsto q(t)$ is known. 
 
At this point the reader will wonder why bothering with introducing the new Hamiltonian $\cH$. The reason is important: $\cH$ is homogeneous of order $2$ in the momenta, and this defines a \textit{geodesic Hamiltonian}. We can write it as 
\be 
\cH = \frac{1}{2} \sum_{A,B=1}^{n+1} g^{AB} p_A p_B \, , 
\ee 
where $p_A = (p_i, p_y)$, $g^{ij} = h^{ij}$, $g^{n+1, n+1} = 2 V$, $g^{n+1, i} = 0$. 

To see what this means let's calculate the associated Lagrangian function: this is given by 
\be \label{eq:Lagrangian_scalar}
\cL = \frac{1}{2} g_{AB} \dot{x}^A \dot{x}^B \, , 
\ee 
where $g_{AB}$ is the inverse matrix of $g^{AB}$, $g_{ij} = h_{ij}$, $g_{n+1, n+1} = \frac{1}{2 V}$, $g_{n+1, i} = 0$, and $x^A = (q^i, y)$.  
 
Suppose that $V$ has a global minimum, then we can assume $V > 0$ anywhere, since $V$ is defined modulo a constant. If $V$ does not have a global minimum then what follows will only apply in a region where $V>0$. We can notice that $g_{n+1,n+1} > 0$ implies that we can write 
\be \label{eq:definition_ds2}
g_{AB} \, dx^A dx^B = ds^2 \, ,   
\ee 
or in other words the left hand side defines a positive quantity that we can think of as the square of the length of an infinitesimal piece of trajectory in the space $(q^i, y)$. $g_{AB}$ then determines how the length is calculated starting from $dq^A$. Technically, $g_{AB}$ is a \textit{Riemannian metric}. Since $g$ is a function of the coordinates in general it will describe a \textit{curved metric}, which means that there will be no set of coordinates such that in a whole neighbourhood of a point the metric can be cast in diagonal form with constant eigenvalues. The square length $ds^2$ as defined by \eqref{eq:definition_ds2} will no longer necessarily be positive in regions where $V$ is allowed to become negative, and we say that the metric changes its signature: we will exclude these regions for now, and in the next section we will show a version of the Eisenhart lift where this problem does not arise. 
 
The equations of motion \eqref{eq:Eisenhart_scalar_eom1}-\eqref{eq:Eisenhart_scalar_eom4}  are equivalent to the Euler-Lagrange equations for $\cL$: 
\be 
\frac{\partial \cL}{\partial x^A} - \frac{d}{dt} \frac{\partial \cL}{\partial \dot{x}^A} = 0 \, , 
\ee 
and it is possible to show that these correspond to finding the extremal trajectories of the functional 
\be \label{eq:functional} 
I[x^A] := \int_{x(s_i)}^{x(s_f)} \sqrt{\frac{1}{2} g_{AB} \frac{dx^A}{ds} \frac{dx^B}{ds}} \, ds 
\ee 
keeping the endpoints fixed. Notice that in the functional above we are using the variable $s$, defined by \eqref{eq:definition_ds2}, to parameterise the trajectory. So what we are seeing is that the original equations of motion correspond to the solution of a geometrical problem: finding the trajectories that extremise \eqref{eq:functional} which represents the total length as measured by the metric $g$.  The Euler-Lagrange equations can be calculated and set in the following form: 
\be 
\frac{d^2 x^A}{ds^2} + \frac{1}{2} g^{AB} \left( \partial_C g_{BD} + \partial_D g_{CB}  - \partial_B g_{CD} \right) \frac{dx^C}{ds} \frac{dx^D}{ds} \, . 
\ee 
This is known as a \textit{geodesic equation} for the metric $g$. Geodesic equations represent motion in a curved space in the absence of any external force. 
 
Now we are ready to show how to include electromagnetic interactions. Given that the scalar potential describing a static electric field can already be represented by $V$, we only need to add a magnetic-type interaction through a vector potential $A_i$. This is done in the standard way modifying eq.\eqref{eq:Hamiltonian_scalar} into 
\be \label{eq:Hamiltonian_full} 
H = \frac{1}{2} \sum_{i,j = 1}^n h^{ij}(q) ( p_i - e A_i ) ( p_j - e A_j) + e^2 V(q) \, . 
\ee 
Following the same steps as before, we lift this Hamiltonian to a homogeneous order 2 Hamiltonian 
\be \label{eq:Hamiltonian_full_lifted} 
\cH = \frac{1}{2} \sum_{i,j = 1}^n h^{ij}(q) ( p_i - p_y A_i ) ( p_j - p_y A_j) + p_y^2 \, V(q) \, , 
\ee
where again we can set $p_y = e$. Following exactly the same steps as before one finds the squared infinitesimal length element given by 
\be 
ds^2 = \sum_{i,j=1,n} h_{ij} dq^i dq^j + \frac{1}{2 V} \left(dy + A_i dq^i \right)^2 \, , 
\ee 
showing that the metric $g$ gets non-diagonal elements $g_{i, n+1} = g_{n+1, i} = \frac{A_i}{2 V}$. In this case too the original equations of motion can be interpreted as geodesic equations for the higher dimensional metric. 
 
To recapitulate, in this section we have seen that we can start from a natural Hamiltonian system and trade its kinetic energy matrix, the scalar and the vector potential for a metric in a higher dimensional space. In this higher dimensional space there is free motion, in the absence of force. We have used the extra dimension to incorporate the potentials in the definition of the metric, the object that defines the geometry.

\subsection{The Lorentzian lift} 
In this section we describe another type of lift that this time involves introducing two extra variables and two conjugated momenta. The lift described here has two main advantages: first, the metric is always well defined unlike in the previous case, and second this type of lift allows to clearly describe geometrically the action of dynamical symmetries of a Hamiltonian system.  
We start again from the Hamiltonian \eqref{eq:Hamiltonian_full}, and this time we build another type of homogeneous, second order Hamiltonian: 
\be \label{eq:Hamiltonian_lorentzian} 
\cH = \frac{1}{2} \sum_{i,j = 1}^n h^{ij}(q) ( \hat{p}_i - p_v A_i ) ( \hat{p}_j - p_v A_j) + p_v^2 \, V(q) + p_u \, p_v \, .  
\ee
This is the same as \eqref{eq:Hamiltonian_full} where we have changed the name of the variable $y$ into $v$, as the variable $v$ appears more frequently in the literature on the subject, and we have added a new variable $u$ and the seemingly innocuous term $p_u p_v$. 
 
Since $\cH$ does not dependend on $u$ or $v$ then Hamilton's equations guarantee that $p_u$ and $p_v$ are constants. We can set $p_v = e$ as before: if we do this then we can see that 
\be 
\cH = H + e p_u \, , 
\ee 
where $H$ is given by \eqref{eq:Hamiltonian_full}. This is not enough yet to recover the original Hamiltonian immediately as in the previous section: we need to do some extra work. What we can do is to restrict our attention from now on only to solutions of the equations of motion for which $\cH = 0$, as $p_u$ can be negative. Then for these solutions on can solve for $p_u$ and find 
\be \label{eq:pu} 
p_u = - \frac{H}{e} \, . 
\ee 
This has the following interpretation: we know from Hamiltonian theory, using Poisson brackets, that $p_u$ generates a translation along the $u$ variable, while $H$ a translation along the time variable $t$ of the original system. Then we can identify 
\be \label{eq:t_def}
t = - \frac{u}{e} \, . 
\ee 
We can use this equation going forward in order to recover the description of a Hamiltonian system with $n$ dimensions, plus the time $t$. Notice that the $q^i$ and $p_i$ equations of motion are, upon fixing $p_v = e$, the same as those for the original Hamiltonian \eqref{eq:Hamiltonian_full}. The equation for $u$ is $\dl{u} = \frac{\partial \cH}{\partial p_u} = p_v = e$, where now we are parameterising trajectories with a parameter $\lambda$ that is not necessarily the time parameter of the original system. In fact, the equation for $u$ allows substituting $u$ instead of $\lambda$ everywhere, and then using \eqref{eq:t_def} we can substitute the time $t$. The equation for $v$ is a bit more involved, but as in the previous section it can be integrated once the functions $q^i(\lambda)$ are known. Therefore, the Hamiltonian \eqref{eq:Hamiltonian_lorentzian} induces, for solutions with $\cH = 0$, the same trajectories of the original Hamiltonian $H$. 
 
We now examine the infinitesimal length element. As in the previous section, we use the trick of reading this from the expression of the Lagrangian. We obtain 
\be \label{eq:Eisenhart_metric_Lorentzian} 
ds^2 = \sum_{i,j=1}^n h_{ij} \, dq^i dq^j + 2 du ( dv - V du + A_i dq^i ) \, . 
\ee 
Here we have indicated, as is customary to do with curved spaces in general, the left hand side with the symbol $ds^2$, however the reader should notice that the right hand side can assume both positive and negative sign. This is actually necessary in order for the condition $\cH = 0$ be possible. The right hand side defines, as before, the metric $g_{AB}$, One could diagonalise $g_{AB}$ and find that all eigenvalues are positive but one, which is negative. This is an example of \textit{Lorentzian metric}; Lorentzian metrics are important for example in the theory of Relativity. It important noticing that the condition $\cH = 0 = \frac{1}{2} g^{AB} p_A p_B$, where in this section $p_A = (p_i, p_v, p_u)$, 
 translates into considering geodesic trajectories that satisfy the equivalent condition $g_{AB} \dl{x}^A \dl{x}^B = 0$: the vector tangent to the trajectory is \textit{null}, and the geodesic a \textit{null geodesic}. 
Differently from the case discussed in the previous section, this metric does not suffer from the issue of being ill-defined in any region. For example its determinant is never zero nor divergent. One of its more interesting features is that it allows describing geometrically the dynamical symmetries of the original system, as we will see in the next section. 
 
One final technical remark: for simplicity we have not allowed the potentials and the matrix $h_{ij}$ to depend explicitly on time, however this can be accommodated in the construction, see for example \cite{GaryPeterPengmingJWMarco2014}.

\section{Geometrisation of dynamical symmetries} 
In this section we will see that the Lorentzian Eisenhart lift allows a geometrical description of dynamical symmetries of the original system. We will describe in detail the chief example of dynamical symmetries of a single free particle as this is a didactical example where everything can be solved in detail, while at the same time retaining a non-trivial structure given by the conformal group of transformations of flat space. We will comment on the extension to Galilean symmetries and to more general higher order dynamical symmetries. 
 
First of all we notice that in general if \eqref{eq:Eisenhart_metric_Lorentzian} is a Lorentzian Eisenhart metric, then also a new metric $\bar{g} = C(u, q^i)^{-1} g$, where $C$ is a function, will be Lorentzian and will share null geodesics with $g$, since the new Hamiltonian will be $\bar{\cH} = C(u, q^i) \cH$, and the equations of motion of this new Hamiltonian only get corrections in the form of new terms of the type $\partial C \cdot \cH$, which vanish for null geodesics. So we are naturally lead to the concept of a \textit{conformal transformation} of the metric: a transformation of configuration space that changes the metric multiplying it times a nonzero factor. Conformal transformations are extremely important in physics, they are relevant for example in condensed matter, particle physics, String Theory, and in some cosmological theories. 
 
It is in fact possible to show that conformal transformations of the Lorentzian lift are in one to one correspondence with Galilean symmetries of the original system, which are time dependent transformations that infinitesimally are given by 
\be \label{eq:Galilean} 
\delta q^i = f^i (q, t) \, , \qquad \delta t = f^t (t) \, , 
\ee 
for a set of functions $(f^i, f^t)$, and such that they act on solutions of the equations of motion transforming them into solutions. These transformations can be lifted to the higher dimensional Lorentzian space, where they generate conformal transformations \cite{GaryPeterPengmingJWMarco2014}. While we do not provide a full proof of this here, we consider an important example to help clarify the relation, that of a free particle. In this case $V=0$, $A_i = 0$ and the original Hamiltonian is  
\be \label{eq:Hamiltonian_free} 
H = \frac{1}{2} \sum_{i,j = 1}^n \delta^{ij}(q) p_i p_j  \, , 
\ee 
where we are considering unit mass. The Eisenhart Lorentzian lift metric is given by 
\be \label{eq:Eisenhart_metric_free} 
ds^2 = \sum_{i,j=1}^n \delta_{ij} \, dq^i dq^j + 2 du  dv \, ,  
\ee  
which is flat space. 
The geodesic Hamiltonian $\cH$ generates trajectories that are straight lines, and is invariant under translations, rotations, time translation. There are also Galilean boosts $q^i \rightarrow q^i + \tilde{p}^i t$, with $\tilde{p}$ constant: these alter the velocity and therefore the energy of the trajectory. However, there are two more Galilean transformations that act on trajectories. These go by the traditional names of dilatations: 
\be \label{eq:Galilean_transformation_d}
\delta q^i = t p^i - \frac{1}{2} q^i \, , \qquad \delta t = - t \, , 
\ee 
and expansions 
\be \label{eq:Galilean_transformation_k}
\delta q^i = t^2 p^i - t q^i \, , \qquad \delta t = - t^2 \, . 
\ee
The reader can explicitly check, using the formulas above, that straight lines $\vec{q}(t) = \vec{q}_0 + \vec{p}_0 t$, with $\vec{q}_0$ and $\vec{p}_0$ constants, are mapped into straight lines: this means that dilatations and expansions are \textit{dynamical symmetries}, they map solutions of the equations of motion into solutions. However, unlike translations, time translations and rotations the new solutions have different energy than the original ones: the Hamiltonian is changed by the dynamical symmetry. For the dilatations one can check that $\delta H = H$, while for expansions $\delta H = 2t H - \vec{q} \cdot \vec{p}$. In fact the $q$ transformations above can be seen to arise from a canonical transformation: one defines phase space functions 
\ba 
D &=& t H - \frac{1}{2} \vec{q} \cdot \vec{p} \, ,   \\ 
K &=& t^2 H - t \vec{q} \cdot \vec{p} + \frac{1}{2} q^2 \, , \label{eq:K} 
\ea 
then the full canonical transformation is generated via Poisson brackets as 
\be \label{eq:canonical_1}
\delta q^i = \{ q^i, C \} \, , \qquad \delta p_i = \{ p_i, C \} \, , 
\ee 
where $C = D, K$. The reader can check that this generates the $q$ part of \eqref{eq:Galilean_transformation_d}, \eqref{eq:Galilean_transformation_k}. 
The functions $H$, $D$, $K$ satisfy, via Poisson brackets, the algebra of the $SL(2, \mathbb{R})$ subgroup of non-relativistic conformal transformations: 
\be 
\{D, H\} = - H \, , \quad \{D, K\} = K \, , \quad \{H, K \} = 2 D \, . 
\ee 
When we add the 3 translations, 3 rotations and 3 boosts, together they form a 12 parameter group called \textit{Schr\"{o}dinger group}, which is in fact the invariance group of the free Schr\"{o}dinger equation \cite{Niederer1972} \footnote{In fact, one should talk about a 13 parameter central extension of the group, the \textit{extended Schr\"{o}dinger group} \cite{GaryDuvalHorvathy1991}.} 
 
The relationship with the Lorentzian Eienhart lift is the following. It is not really restrictive to work with $e=1$ in this section since the potentials are zero. Then, using eqs.\eqref{eq:pu}, \eqref{eq:t_def} we can lift the quantities $D$ and $K$ to new higher dimensional quantities 
\ba 
\hat{D} &=& u p_u + \frac{1}{2} \sum_{i=1}^n q^i \hat{p}_i \, , \nn \\ 
\hat{K} &=& - u^2 p_u - u \sum_{i=1}^n q^i \hat{p}_i + \frac{1}{2} q^2 p_v \, .  \nn 
\ea 
To obtain these equations we have followed two rules: first, we added a factor of $p_v$ to the last term in the definition of $K$, eq.\eqref{eq:K}, so that both $\hat{D}$ and $\hat{K}$ are homogeneous in momenta, and second we noticed that since $t = - \lambda$, with $\lambda$ the parameter for higher dimensional trajectories, then a sign factor arises between the original momenta $p_i$ and higher dimensional momenta $\hat{p}_i$. One can then see that $\hat{D}$ generates, via Poisson brackets and with a recipe analogous to \eqref{eq:canonical_1}, the transformation 
\be 
\delta u = u \, , \quad \delta v = 0 \, , \quad \delta q^i = \frac{q^i}{2} \, , 
\ee 
under which the higher dimensional metric \eqref{eq:Eisenhart_metric_free} transforms as $\delta (ds^2) = ds^2$: this is a conformal transformation. Similarly, $\hat{K}$ generates the transformation 
\be 
\delta u = - u^2 \, , \quad \delta v = \frac{q^2}{2} \, , \quad \delta q^i = - u q^i \, , 
\ee 
under which $\delta (ds^2) = - 2 u ds^2$. This is an example of the more general fact discussed in \cite{GaryPeterPengmingJWMarco2014}: conformal transformations of the Eisenhart lift \eqref{eq:Eisenhart_metric_free} are in one to one correspondence with Galilean transformations \eqref{eq:Galilean} as long as they transform solutions into solutions. This kind of dynamical symmetries of the original system therefore have a geometrical interpretation in the higher dimensional space. We conclude mentioning that also dynamical symmetries of higher order in the momenta can be described using the lift. In this case, however, they do not admit a purely geometrical description, but rather they correspond to transformations in phase space \cite{GaryDavidClaudeHouri2011,AntonGalajinsky,MarcoReview}. An example of lifting the Runge-Lenz vector of the Kepler problem can be found in \cite{GaryDuvalHorvathy1991}.

\section{Some remarks on modern directions of research} 
As we have seen in the previous sections, the Eisenhart lift can be used as a tool to didactically introduce advanced geometrical ideas using mathematics that is available to Physics undergraduates towards the end of their studies and master degree students. One more reason that makes the lift appealing is the fact that it is being actively studied in modern research, and in this section we will provide short examples and references to the literature. Some of the concepts used in this section are more advanced than in the rest of the work, such as covariant derivatives and curvature, however it is not mandatory to know these in order to understand the underlying ideas and how the Eisenhart lift can be useful. 
 
One of the areas where the lift has been used is to produce non-trivial examples of Killing and conformal Killing tensors of rank strictly higher than two. Killing tensors are symmetric objects $K_{A_1 \dots A_p}$ that satisfy 
\be 
\nabla_{(A_1} K_{A_2 \dots A_{p+1})} = 0 \, , 
\ee 
where $\nabla$ is the Levi-Civita covariant derivative built from a metric. Such tensors generate conserved quantities for the geodesic motion and also play a part in the integrability of other types of physical equations in curved space. Examples with $p=1,2$ are known since long, however the case $p\ge 3$ had proven elusive. It has been recently possible to build such tensors by considering Hamiltonian systems with conserved quantities of higher order in the momenta, of which there are known examples, and lifting these to higher dimensions using the Eisenhart lift, thus obtaining conserved quantities for geodesic motion that are of higher order in the momenta. For examples the reader can see \cite{GaryDavidClaudeHouri2011,AntonGalajinsky,GaryMarcoToda}. 
 
Another important area of study is the description of chaotic features of dynamical systems using the geometrical properties of the Eisenhart lift metric. One can obtain a local description of how neighbouring geodesics evolve by writing an equation for perturbed geodesics: 
\be 
q^A(\lambda) = q_0^A(\lambda) + J^A(\lambda) \, , 
\ee 
where $q_0$ represents a geodesic and $J^A$ a perturbation. Then the evolution of $J$ is given by 
\be 
\frac{\nabla^2 J^A}{d\lambda^2} + R^A_{BCD} \frac{dq_0^A}{d\lambda} J^B \frac{dq_0^C}{d\lambda} = 0 \, , 
\ee 
where $\frac{\nabla }{d\lambda}$ is the covariant derivative along the flow of $q_0$ and $R^A_{BCD}$ the Riemann curvature tensor. This equation provides a link between the curvature of space and Lyapunov exponents, that enter in the description of how nearby trajectories separate. Analysing this equation is not easy and there is not yet a definite answer to what is the exact geometrical feature associated to chaos. There are indications that curvature fluctuations play an important role in this. We refer the reader to \cite{PettiniBook} and references therein for more details. 
 
The lift has also been used to describe dualities between different theories \cite{GaryDuvalHorvathy1991,GaryPeterPengmingJWMarco2014}, and a generalised lift has been introduced in \cite{GaryMarcoToda} to describe symmetries that relate systems with different values of the coupling constants. However, arguably one of the more interesting modern applications of the lift is that to non-relativistic holography. Holography is a concept that relates a quantum field theory in $d$ dimensions to a theory of quantum gravity, described in terms of strings, in a higher dimensional space, which conventionally is called the \textit{bulk}. In particular, symmetries of the field theory correspond to symmetries of the theory in the bulk. The holographic description is at the same time very useful and difficult to prove exactly since it tends to relate the regime where one of the two theories is weakly coupled to that where the other is strongly coupled. For relativistic theories the bulk has normally one dimension more than the field theory space, however in the case of non-relativistic theories it seems that two extra dimensions are needed. In fact, this happens because the Lorentzian Eisenhart lift makes its appearence. For example, one metric that has been often used to describe the bulk is given by 
\be 
\bar{ds}^2 = \frac{1}{r^2} \left[ \sum_{i=1}^n dq^i dq^i + 2 du dv  - \frac{du^2}{r^2} + dr^2 \right] \, . 
\ee 
One can recognise here the metric \eqref{eq:Eisenhart_metric_Lorentzian}, multiplied times a factor of $r^{-2}$, where $r$ is a new bulk coordinate, and including in $\bar{ds}^2$ a term for $dr^2$. It is possible to show that the isometries of the bulk metric correspond to conformal transformations of the Eisenhart metric that is written inside the square brackets. 
The metric above is a solution of Einstein's equations of motion that represents a travelling gravitational wave.

\section{Conclusions\label{sec:conclusions}} 
In this work we have given a pedagogical presentation of the Eisenhart lift that only requires knowledge of Hamiltonian dynamics as a prerequisite. We have shown how this leads naturally to concepts of curved spaces, Riemannian and Lorentzian, to extra dimensions and geometrisation of interactions, and we have also given an introduction to the fact that (lower order in momenta) dynamical symmetries can be described geometrically via conformal transformations in higher dimensional space. We believe that this type of presentation offers an opportunity to undergraduate as well as some master students to learn about modern geometrical concepts and techniques without requiring heavy mathematics or long introductions to modern physical theories. We have also provided a brief description of recent research directions that show the Eisenhart lift is a subject of current interest that can be applied to several different physical problems.

\vspace{0.2cm}

\section*{Acknowledgments} 
\noindent M. Cariglia is funded by Universidade Federal de Ouro Preto under the project "Simetrias escondidas de sistemas unidimensionais de muitos corpos". F. Kelmer Alves acknowledges financial support under the program 'Inicia\c c\~ao Cient\'ifica' PROBIC/FAPEMIG/UFOP 12/2013. 



\begin{thebibliography}{10}


\bibitem{Kaluza} 
T. Kaluza, \textit{Zum Unit\"{a}tsproblem in der Physik}, Sitzungsber. Preuss. Akad. Wiss, Berlin 966–972 (1921). 
 
\bibitem{Klein} 
O. Klein, \textit{Quantentheorie und f\"{u}nfdimensionale Relativit\"{a}tstheorie}, Zeitschr. Phys. \textbf{37}(12) 895-906 (1926). 

\bibitem{Eisenhart1928} 
L. P. Eisenhart, \textit{Dynamical trajectories and geodesics}, Annals. Math. \textbf{30} 591-606 (1928). 
 
\bibitem{GaryPeterPengmingJWMarco2014} 
M. Cariglia, G. W. Gibbons, J. W. van Holten, P. A. Horv\'athy and P. M. Zhang, \textit{Conformal Killing tensors and covariant Hamiltonian dynamics} (2014), arXiv:1404.3422. 
 
\bibitem{Niederer1972} 
U. H. Niederer, \textit{The maximal kinematical invariance groups of Schr\"{o}dinger equations with arbitrary potentials}, Helv. Phys. Act. \textbf{45} 802-810 (1972) 
 
\bibitem{GaryDuvalHorvathy1991} 
C. Duval, G. Gibbons and P. Horv\'athy, \textit{Celestial mechanics, conformal structures, and gravitational waves}, Phys. Rev. D \textbf{43} (12) 3907-3922 (1991). 
 
\bibitem{GaryDavidClaudeHouri2011}  
G. Gibbons, T. Houri, D. Kubiz\v{n}\'ak, and C. M. Warnick, \textit{Some spacetimes with higher rank Killing-St\"{a}ckel tensors}, Phys. Lett. \textbf{B 700}(1) 68-74 (2011). 
 
\bibitem{AntonGalajinsky}
A. Galajinsky, \textit{Higher rank Killing tensors and Calogero model}, Phys.\ Rev.\ D  \textbf{85} 085002 (2012). 
 
\bibitem{MarcoReview} 
M. Cariglia, \textit{Hidden Symmetries of Dynamics in Classical and Quantum Physics}, to be published in Rev. Mod. Phys. (2014). 
 
\bibitem{GaryMarcoToda} 
M. Cariglia, G. Gibbons, \textit{Generalised Eisenhart lift of the Toda chain}, J. Math. Phys. \textbf{55} 022701 (2014). 
  
\bibitem{PettiniBook} 
M. Pettini, \textit{Geometry and Topology in Hamiltonian Dynamics and Statistical Mechanics}, Springer, New York 2007. 
 

 





\end{thebibliography}

\providecommand{\href}[2]{#2}\begingroup\raggedright\endgroup

\end{document}